# Towards interpretable prediction of recurrence risk in breast cancer using pathology foundation models

Jakub R. Kaczmarzyk[1,2,3,*], Sarah C. Van Alsten[4,5], Alyssa J. Cozzo[4], Rajarsi Gupta[1], Peter K. Koo[2], Melissa A. Troester[4,5,6], Katherine A. Hoadley[4,7,†], Joel H. Saltz[1,†]

[1]Department of Biomedical Informatics, Stony Brook University, Stony Brook, NY, 11794, USA
[2]Simons Center for Quantitative Biology, Cold Spring Harbor Laboratory, Cold Spring Harbor, NY, 11724, USA
[3]Medical Scientist Training Program, Stony Brook University, Stony Brook, NY, 11794, USA
[4]Lineberger Comprehensive Cancer Center, University of North Carolina at Chapel Hill, Chapel Hill, NC, 27599, USA
[5]Department of Epidemiology, University of North Carolina at Chapel Hill, Chapel Hill, NC, 27599, USA
[6]Department of Pathology and Laboratory Medicine, School of Medicine, University of North Carolina at Chapel Hill, Chapel Hill, NC, 27599, USA
[7]Department of Genetics, University of North Carolina at Chapel Hill, Chapel Hill, NC, 27599, USA

*jakub.kaczmarzyk@stonybrookmedicine.edu
[†]KAH and JHS jointly supervised the work.

## Abstract

Transcriptomic assays such as the PAM50-based ROR-P score guide recurrence risk stratification in non-metastatic, ER-positive, HER2-negative breast cancer but are not universally accessible. Histopathology is routinely available and may offer a scalable alternative. We introduce MAKO, a benchmarking framework evaluating 12 pathology foundation models and two non-pathology baselines for predicting ROR-P scores from H&E-stained whole slide images using attention-based multiple instance learning. Models were trained and validated on the Carolina Breast Cancer Study and externally tested on TCGA BRCA. Several foundation models outperformed baselines across classification, regression, and survival tasks. CONCH achieved the highest ROC AUC, while H-optimus-0 and Virchow2 showed top correlation with continuous ROR-P scores. All pathology models stratified CBCS participants by recurrence similarly to transcriptomic ROR-P. Tumor regions were necessary and sufficient for high-risk predictions, and we identified candidate tissue biomarkers of recurrence. These results highlight the promise of interpretable, histology-based risk models in precision oncology.

# Introduction

Hormone receptor (HR)-positive, HER2-negative breast cancers account for over 70% of all breast cancer cases and carry a substantial risk of long-term recurrence.[1–3] A meta-analysis of 62,923 estrogen receptor (ER)-positive women reported 20-year recurrence risks ranging from 10% to 41%,[4] and a Danish cohort study with over 30 years of follow-up found recurrence rates of 13.5% to 34.3% among ER-positive patients.[5] While patients at high risk of recurrence benefit most from adjuvant chemotherapy, accurately identifying those at low risk is equally important to avoid unnecessary treatment and its associated side effects.[6–11]

To address this clinical need, transcriptomic assays like the PAM50-based risk of recurrence (ROR-P) score have been clinically validated for the prediction of recurrence risk in ER-positive, HER2-negative patients.[12–20] However, these assays are not universally available and may potentially delay decision-making due to turnaround times of several days to week.[21] Given the increasing digitization of histopathology, artificial intelligence (AI) applied to hematoxylin-and-eosin (H&E)-stained whole slide images (WSIs) offers a promising and scalable alternative for biomarker inference, particularly in settings where transcriptomic testing is inaccessible. Previous studies have demonstrated that AI models can infer HR status,[22–25] PAM50 molecular subtypes,[23,26] and even ROR-P[23,27,28] directly from H&E-stained WSIs.

While encouraging, most prior work has relied on task-specific models or feature extractors pretrained on natural images, such as those from ImageNet.[23,27–29] These models may not optimally capture the morphological complexity of histopathology. In contrast, recent advances in general-purpose, pretrained pathology foundation models have demonstrated strong performance across diverse WSI-level tasks.[30–36] However, their utility for ROR-P prediction and ability to predict long-term recurrence in breast cancer have not been systematically evaluated. Benchmarking these

pathology foundation models for risk prediction could accelerate progress in tissue-based prognostics.[37]

Moreover, despite the widespread use of attention-based multiple instance learning (ABMIL) for WSI-level prediction tasks,[38,39] the interpretability of these ABMIL models remains poorly understood.[40] For clinical deployment, it is essential not only to assess predictive performance but also to understand how models arrive at their predictions. Interpretability methods can help determine rely on histologically meaningful features or spurious correlations, and may aid in identifying potential biases, failure models, or novel biomarkers. While attention weights from ABMIL are widely used for interpretation, they are best considered as tools for generating hypotheses about which tissue regions may be informative. Attention highlights areas the model may be focusing on but does not necessarily reveal its true decision-making process.[41] In practice, attention can be unreliable, especially in the presence of redundant or correlated features. To move from hypothesis generation to hypothesis testing, perturbation-based methods offer a more rigorous framework by systematically evaluating how altering specific tissue regions affects model outputs.[40] Despite their potential, no prior studies to our knowledge have applied such virtual experiments to assess the interpretability of foundation models for breast cancer recurrence risk stratification.

To address these gaps, we developed **MAKO** (Mammary Analysis for Knowledge of Outcomes), a comprehensive benchmarking framework for inferring recurrence risk from H&E-stained WSIs in early breast cancer (**Figure 1**). MAKO evaluates 14 pretrained feature extractors using ABMIL, including 12 pathology-specific foundation models and two general-purpose vision encoders. Models were trained using data from the Carolina Breast Cancer Study (CBCS), a large, diverse cohort with long-term recurrence follow-up, and externally validated on The Cancer

Genome Atlas Invasive Breast Carcinoma dataset (TCGA BRCA).[42] We focused specifically on predicting the PAM50-based ROR-P score, comparing models trained for classification of risk categories (i.e., low/medium vs. high) with those trained to regress the continuous ROR-P score, the latter of which may better preserve underlying biological variation.[43] In addition to benchmarking predictive performance, we assessed model interpretability by examining the contribution of tumor epithelium versus surrounding tissue to the predictions. We conducted further virtual experiments in tissue to identify histologic features associated with high recurrence risk, advancing efforts toward biologically grounded and clinically actionable AI in computational pathology. Together, these efforts provide a comprehensive evaluation of pathology foundation models for breast cancer risk stratification and establish MAKO as a resource for benchmarking and discovery.

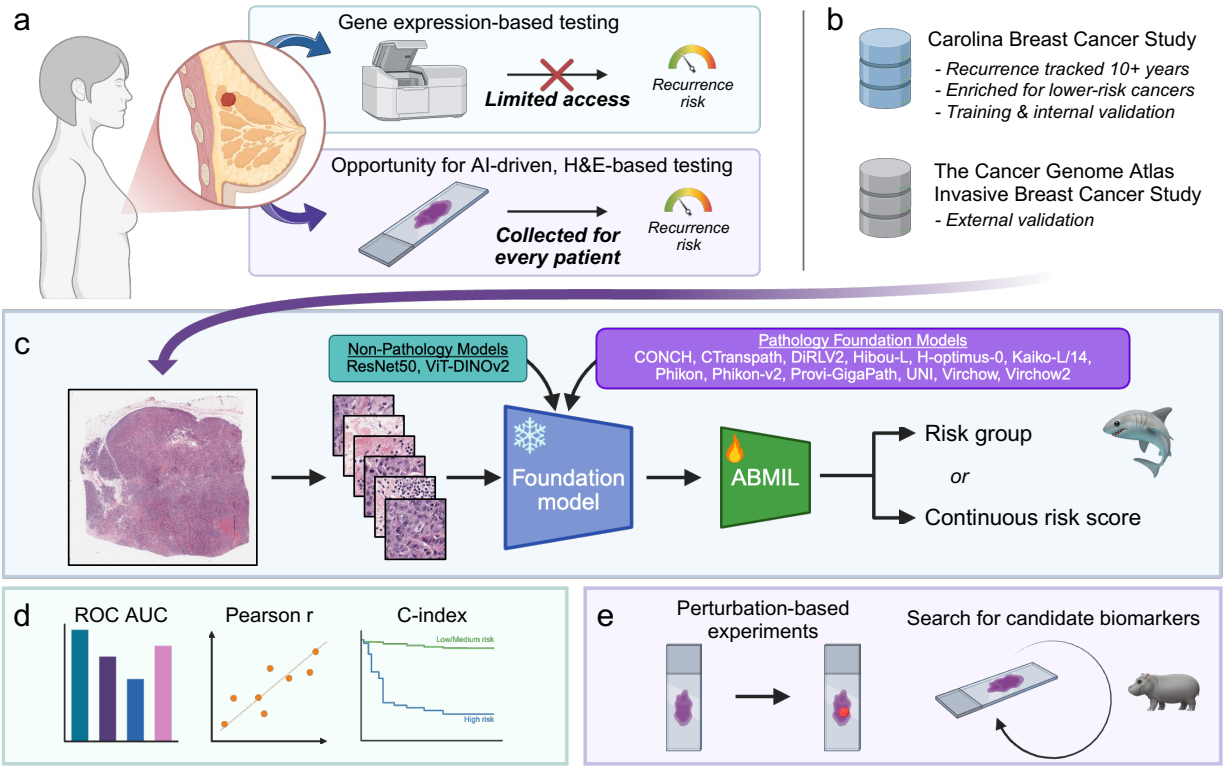

Figure 1. **Overview of MAKO. a**, Gene expression assays such as the ROR-P score provide recurrence risk estimates in ER-positive, HER2-negative breast cancer, but there are limitations to their universal use. In contrast, H&E-stained slides are collected routinely for every patient with breast cancer and offer an accessible, underused data source for AI-based risk prediction. Recent pathology foundation models have not been benchmarked for recurrence risk prediction. We explore this in the present study. **b**, The Carolina Breast Cancer Study (CBCS) was used for model development and internal validation. CBCS is a cohort with high-fidelity longitudinal follow-up, enabling robust evaluation of recurrence prediction. External validation was performed using The Cancer Genome Atlas Invasive Breast Cancer Study (TCGA BRCA), which includes H&E whole slide images (WSIs) and gene expression data, though recurrence tracking is less complete. **c**, WSIs were divided into fixed-size image patches, which were then embedded using one of 14 pretrained encoders, including 12 pathology foundation models and two baseline models. Patch embeddings were aggregated using attention-based multiple instance learning (ABMIL) to produce a WSI-level prediction. For each pretrained encoder, two models were evaluated: a classification model discriminating between low/medium and high ROR-P, as well as a regression model directly predicting the continuous ROR-P score. **d**, ROC AUC was used to evaluate classification of ROR-P risk groups, while Pearson correlation coefficient (r) was used to measure performance of regression models. Concordance index (C-index) was used to assess time-to-recurrence prediction in survival analysis, using model predictions. **e**, We applied HIPPO to perform virtual experiments on WSIs to determine whether tumor regions were necessary and sufficient for high-risk predictions. To discover candidate tissue biomarkers of recurrence risk, we used HIPPO to identify the smallest regions of tissue capable of converting a low-risk prediction into a high-risk prediction when inserted into a different slide.

# Results

Using attention-based multiple instance learning (ABMIL), we evaluated 12 foundation models pretrained on histopathology and compared their performance to two non-pathology baselines: ResNet50 pretrained on ImageNet and ViT-DINOv2, a vision transformer trained on 142 million natural images using self-supervised learning. This design enabled assessment of the added value of pathology-specific pretraining. Models were evaluated on three tasks: classification of ROR-P risk categories (low/medium vs. high), prediction of continuous ROR-P scores, and stratification by time to recurrence. Training and internal validation were performed on the Carolina Breast Cancer Study (CBCS) using 10-fold cross validation, with external validation on TCGA BRCA (**Table 1**). While training included all patients, analyses focused on ER-positive, HER2-negative tumors, the target population for clinical ROR-P testing.

In addition to benchmarking performance, we applied the HIPPO framework to identify histologic features used by the models. We evaluated whether tumor regions were necessary and sufficient for accurate ROR-P prediction, and we identified regions sufficient to predict high-risk scores. These analyses offer insight into the spatial patterns driving model predictions and their potential relevance as interpretable biomarkers.

Table 1. Characteristics of cohorts.

| | CBCS (N=1339) | TCGA BRCA (N=1050) |
|---|---|---|
| **Age (years)** | | |
| Mean (SD) | 51.5 (11.2) | 58.6 (13.2) |
| Median [Min, Max] | 49.0 [24.0, 74.0] | 59.0 [26.0, 89.0] |
| Missing | 0 (0%) | 15 (1.4%) |
| **Race** | | |
| Black | 696 (52.0%) | 167 (15.9%) |
| Non-Black | 643 (48.0%) | 787 (75.0%) |
| Missing | 0 (0%) | 96 (9.1%) |
| **Receptor Status** | | |
| ER+/HER2– | 868 (64.8%) | 642 (61.1%) |
| Not ER+/HER2– | 471 (35.2%) | 372 (35.4%) |
| Missing | 0 (0%) | 36 (3.4%) |
| **Tumor Size (mm)** | | |
| Mean (SD) | 26.7 (22.0) | — |
| Median [Min, Max] | 21.0 [2.00, 200] | — |
| Missing | 1 (0.1%) | 1050 (100%) |
| **Node Status** | | |
| Negative | 766 (57.2%) | 499 (47.5%) |
| Positive | 570 (42.6%) | 530 (50.5%) |
| Missing | 3 (0.2%) | 21 (2.0%) |
| **TNM Stage** | | |
| Stage I | 506 (37.8%) | 173 (16.5%) |
| Stage II | 592 (44.2%) | 597 (56.9%) |
| Stage III | 205 (15.3%) | 236 (22.5%) |
| Stage IV | 36 (2.7%) | 19 (1.8%) |
| Missing | 0 (0%) | 25 (2.4%) |
| **Intrinsic Subtype** | | |
| Luminal A | 519 (38.8%) | 566 (53.9%) |
| Luminal B | 274 (20.5%) | 215 (20.5%) |
| HER2-enriched | 149 (11.1%) | 81 (7.7%) |
| Basal | 397 (29.6%) | 188 (17.9%) |
| **ROR-P Group** | | |
| Low | 242 (18.1%) | 278 (26.5%) |
| Medium | 657 (49.1%) | 554 (52.8%) |
| High | 440 (32.9%) | 218 (20.8%) |

**Benchmarking ROR-P group classification**

Across all models evaluated, pathology foundation models consistently outperformed the ResNet50 baseline for classification of ROR-P risk groups in the CBCS dataset, as measured by ROC AUC. The top-performing model, CONCH, achieved an AUC of 0.809, representing an 8.6% relative improvement. Six additional encoders also had statistically significant improvements over ResNet50 after multiple testing correction included UNI, Phikon, CTransPath, Phikon-v2, Virchow, and DiRLv2. CONCH was the only model to reach a significantly higher ROC AUC than ViT-DINOv2 (**Figure 2a,b**, **Supplementary Table 1**).

In the TCGA BRCA cohort of patients with non-metastatic ER-positive, HER2-negative tumors, only two categorical models demonstrated statistically significant improvements in ROC AUC over the ResNet50 baseline. The CONCH model achieved the highest AUC at 0.852, corresponding to a 10.3% relative improvement. CTransPath also achieved a significantly higher AUC of 0.829. Six additional models (i.e., Virchow2, H-optimus-0, Phikon-v2, UNI, Phikon, and Prov-GigaPath) showed numerically higher ROC AUCs compared to ResNet50, but these did not reach statistical significance after adjustment for multiple comparisons (**Figure 2c,d**, **Supplementary Table 1**).

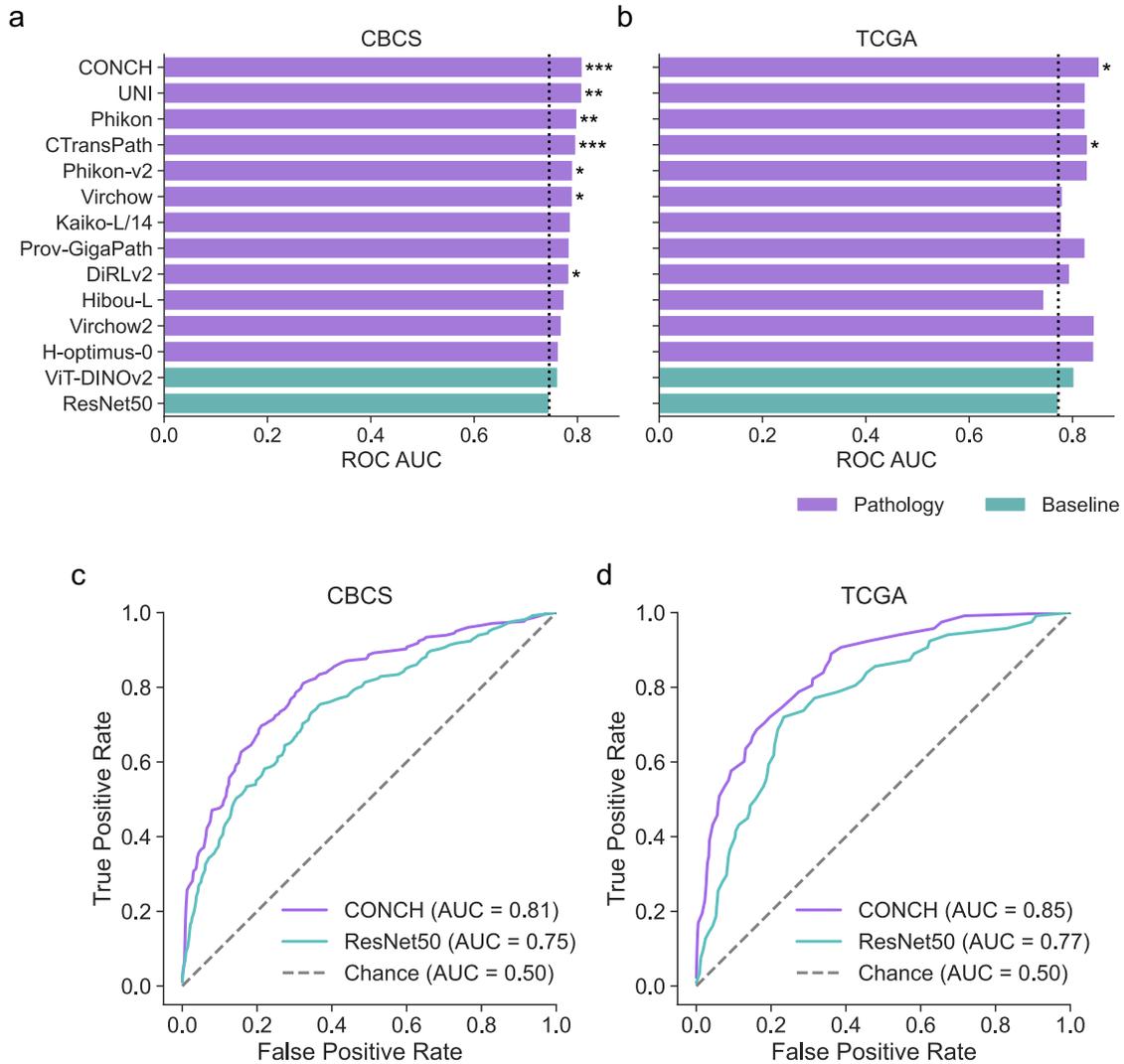

Figure 2. **Benchmarking ROR-P group classification. a, b**, bar plots of receiver operating characteristic area under the curve (ROC AUC) for predicting low/medium versus high ROR-P in ER-positive, HER2-negative participants, as evaluated in **(a)** CBCS (internal) and **(b)** TCGA BRCA (external) cohorts. High ROR-P was treated as the positive category. Models are sorted by ROC AUC in CBCS, with pathology-specific foundation models (purple) compared against ResNet50 and ViT-DINOv2 baselines (teal). Asterisks indicate a statistically significant improvement in ROC AUC compared to ResNet50 within each dataset (DeLong test with FDR correction; $P_{adj} < 0.05$ *, $P_{adj} < 0.01$ **, $P_{adj} < 0.001$ ***). **c, d**, ROC curves from the ER-positive, HER2-negative cohort of **(c)** CBCS and **(d)** TCGA BRCA, illustrating the performance of CONCH (purple) and ResNet50 (teal). Chance performance is shown as a dashed line.

**Benchmarking ROR-P score regression**

In patients with non-metastatic, ER-positive, HER2-negative breast cancer in the CBCS cohort, eleven of twelve pathology foundation models significantly outperformed the ResNet50 baseline in predicting continuous ROR-P scores, as measured by Pearson correlation with the true ROR-P. The ResNet50 model achieved a baseline correlation of 0.541, and the H-optimus-0 encoder achieved the strongest performance, with a correlation of 0.638 (**Figure 3a,b**, **Supplementary Table 2**).

Despite being the highest performer in CBCS, H-optimus-0 did not generalize as well in the TCGA cohort of patients with ER-positive, HER2-negative breast cancer, achieving a lower correlation with true ROR-P scores compared to ResNet50. None of the models evaluated in TCGA demonstrated statistically significant improvements in Pearson correlation relative to ResNet50 after multiple testing correction. Virchow2 and CTransPath showed the strongest numerical improvements, but their adjusted p values narrowly exceeded the significance threshold. Several encoders that performed well in CBCS, including CONCH, Prov-GigaPath, and UNI, exhibited modest gains in correlation, while others such as Phikon, Kaiko-L/14, and H-optimus-0 showed decreased performance (**Figure 3c,d**). These results suggest limited cross-cohort generalizability of continuous ROR-P models trained on digital histopathology.

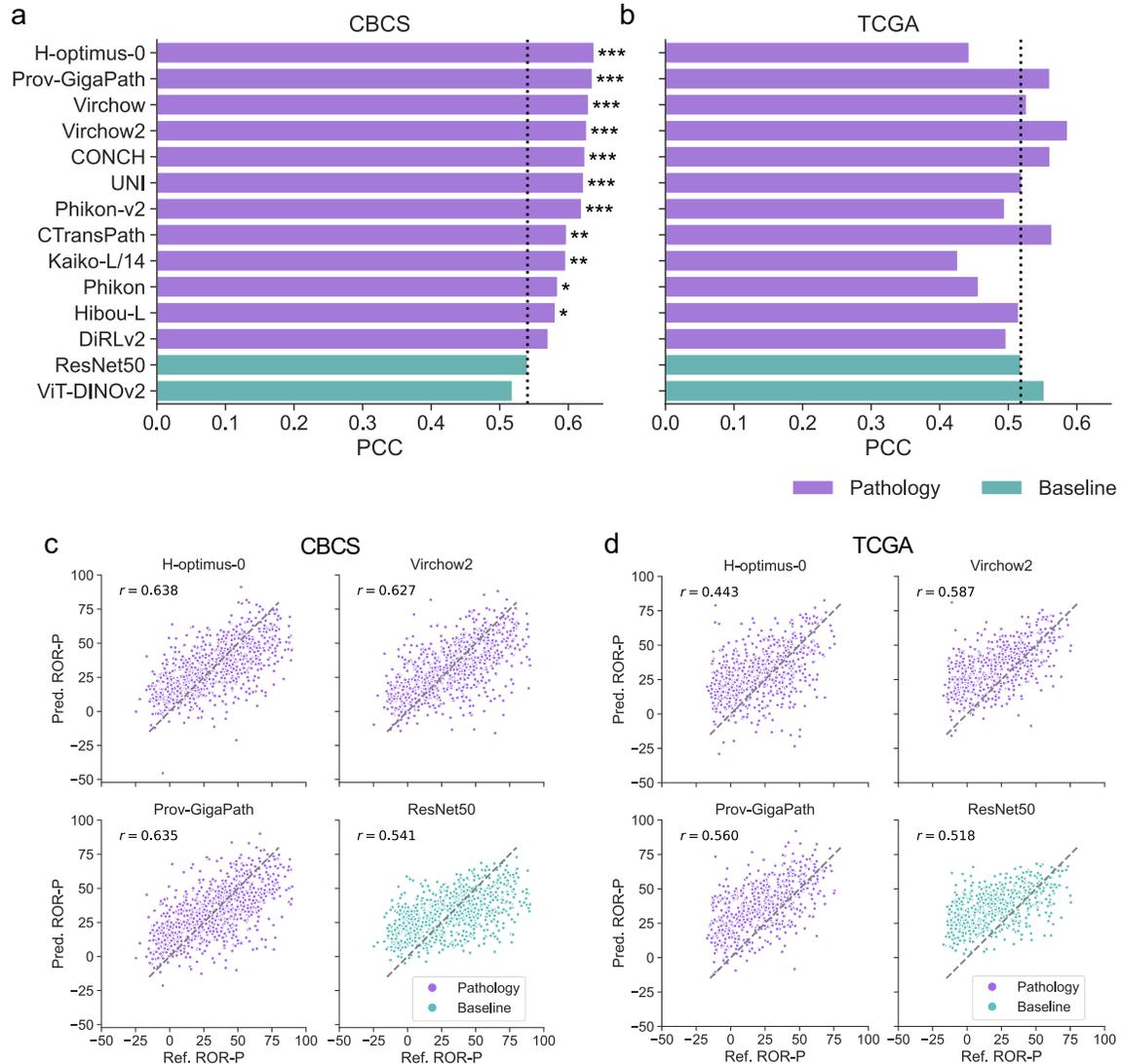

Figure 3. **Benchmarking ROR-P score regression. a, b**, Bar plots of Pearson correlation coefficient (PCC, $r$) for predicting continuous ROR-P scores in ER-positive, HER2-negative participants, as evaluated in **(a)** CBCS (n=883 WSIs, 847 participants) and **(b)** TCGA BRCA (n=628 WSIs, 628 participants) cohorts. Models are sorted by PCC in CBCS, with pathology-specific foundation models (purple) compared against ResNet50 and ViT-DINOv2 baselines (teal). Asterisks indicate a statistically significant improvement in PCC compared to ResNet50 within each dataset (Meng's Z test with FDR correction; $P_{adj} < 0.05$ *, $P_{adj} < 0.01$ **, $P_{adj} < 0.001$ ***). **c, d**, Representative scatter plots of reference transcriptomic ROR-P predicted ROR-P in **(c)** CBCS and **(d)** TCGA BRCA, illustrating the performance of pathology foundation models (purple) and baseline models (teal). Each point represents results from one WSI. The identity function is shown as a dashed line.

**Benchmarking prediction of recurrence events**

We evaluated histology-based models for their ability to predict actual recurrence events using the concordance index (C-index, $C$) as the primary performance metric. Among 847 participants with ER-positive, HER2-negative breast cancer participants in the CBCS cohort, 107 experienced a recurrence event within 10 years of study enrollment. We first evaluated models trained to classify binarized ROR-P risk groups by comparing their prognostic performance to that of the transcriptomic ROR-P assay. All ABMIL-based univariate Cox models demonstrated statistically significant stratification of patients based on recurrence (all $P_{adj} < 0.05$, log-rank test, FDR corrected), except for the H-optimus-0 model ($P = 8.66 \times 10^{-2}$, log-rank test, FDR corrected). In addition, the C indices from ABMIL model predictions were compared to the C index from the transcriptomic assay using the statistical method proposed by Kang et al.[44] All ABMIL models achieved C indices that were not inferior to the transcriptomic assay (all $P_{adj} > 0.05$, Kang et al. test, FDR corrected), indicating comparable performance in risk stratification within this cohort (**Figure 4a,d**). None of the ABMIL models demonstrated statistically significant differences in C-index compared to the ResNet50 baseline (all $P_{adj} > 0.05$, Kang et al. test, FDR corrected).

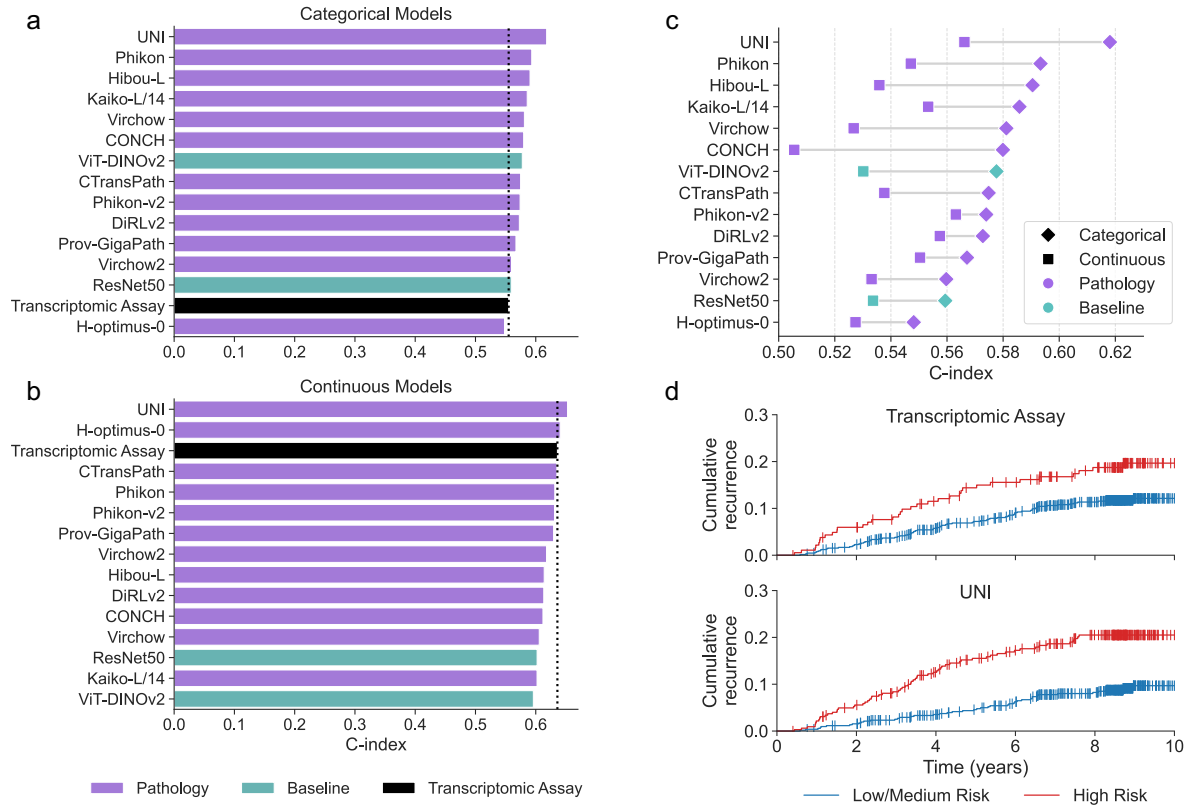

Figure 4. **Benchmarking pathology foundation models for predicting recurrence events. a,** Concordance index (C-index) of models trained to classify ROR-P risk groups, evaluated using recurrence-free survival in the Carolina Breast Cancer Study (CBCS). The transcriptomic ROR-P assay is shown in black as the reference standard. Models are sorted by C-index. **b**, C-index values for models trained to predict continuous ROR-P scores, with reference standard shown in black. Models are sorted by C-index. **c**, Dumbbell plot comparing C-indexes achieved by categorical models (diamonds) and thresholded continuous predictions (squares) for each model. Categorical models consistently outperformed their thresholded continuous counterparts. **d**, Cumulative recurrence curves stratified by low/medium (blue) vs. high risk (red) groups. Predictions from the transcriptomic ROR-P assay and from the categorical UNI model both showed clear separation, with the UNI model achieving comparable prognostic stratification.

Next, we benchmarked models trained to predict continuous ROR-P scores using the C-index. The continuous prediction scores from all ABMIL models were significantly associated with recurrence-free survival (all $P_{adj} < 0.05$, Wald test, FDR corrected), indicating that higher scores corresponded to increased risk of recurrence. Similar to the categorical models, no continuous ABMIL-based Cox model differed significantly from the transcriptomic ROR-P assay

(all $P_{adj} > 0.05$, Kang et al. test, FDR corrected) or from the ResNet50 baseline model (all $P_{adj} > 0.05$, Kang et al. test, FDR correction) (**Figure 4b**).

We observed that C indices were generally higher for models trained to predict continuous ROR-P scores than for those trained to classify binarized ROR-P risk groups (**Figure 4a,b**). To determine whether this performance gap reflected genuine model differences or was instead due to differences in resolution, we applied the same threshold used by the transcriptomic ROR-P assay to the predicted continuous scores. When evaluated in a directly comparable classification framework, the categorical models consistently achieved higher C indices than their thresholded continuous counterparts (**Figure 4c**), but after correction for multiple comparisons, only CONCH achieved a significantly higher C index in the categorical setting ($P = 1.61 \times 10^{-2}$, Kang et al. test, FDR corrected). These findings suggest that while continuous models benefit from finer granularity in survival modeling, categorical models trained explicitly to align with clinical thresholds may better capture risk group distinctions relevant to treatment decision-making. It is also possible that the transcriptome thresholds are not optimized for the foundation model predictions.

Due to limited clinical follow-up data in TCGA,[45] this cohort was used for secondary analysis, while CBCS remained the primary dataset for evaluating recurrence prediction. In the ER-positive, HER2-negative, non-metastatic cohort of TCGA BRCA ($n = 628$), 49 participants experienced a recurrence event within 10 years of study enrollment. In this subset of TCGA BRCA, the transcriptomic ROR-P assay did not significantly stratify participants by recurrence when binarized ($C = 0.533$, $P = 1.09 \times 10^{-1}$, log-rank test) or used as a continuous score ($C = 0.468$, $P = 4.67 \times 10^{-1}$, Wald test). Likewise, none of the ABMIL models achieved significant stratification in TCGA BRCA (all $P_{adj} > 0.05$, log-rank test, FDR corrected).

**Attention is insufficient for interpretation**

With ABMIL models, each patch is assigned a weight reflecting its contribution to the model's prediction.[38] To better understand how these models inferred recurrence risk, we qualitatively analyzed high-attention patches across specimens stratified by predicted ROR-P groups. In high ROR-P predictions, the most attended patches frequently captured nuclear pleomorphism, disordered architecture, and tumor–stroma interfaces (**Supplementary Figure 2a**), whereas in low/medium ROR-P predictions, high-attention patches were often localized to stromal regions (**Supplementary Figure 2b**). In general, attention in high-risk cases was concentrated within tumor epithelial regions, suggesting that the model prioritizes tumor morphology in its high-risk assessments **(Supplementary Figure 2c)**. However, we also identified specimens in which attention was diffuse and without clear focus **(Supplementary Figure 2d)**. These ambiguous attention maps limited interpretability and raised questions about which tissue regions were actually driving model predictions.

**Tumor regions are necessary and sufficient for high ROR-P predictions**

To address these limitations and rigorously quantify the contribution of tumor tissue, we used HIPPO, a perturbation-based explainability framework specifically designed to assess how localized tissue regions influence predictions in weakly-supervised models, like those developed in the present study.[40] While attention maps offer indirect insights, they are not guaranteed to reflect causal model behavior.[41] Perturbation-based approaches, by contrast, enable systematic testing of tissue importance through controlled input modification. Motivated by the biology of the transcriptomic ROR-P assay (which was developed using genes that were intrinsic to tumor found in repeated sampling of tumor tissue[12,19]), we hypothesized that tumor would be both necessary and sufficient for high ROR-P predictions by our models. Using HIPPO, we generated synthetic slides by removing tumor patches and observed the resulting impact on ROR-P predictions. Across all models, removal of tumor regions led to a significant reduction in high-risk scores ($P < 0.05$, two-sided paired t-test), with effect sizes ranging from $-1.77$ to $-0.12$ (Cohen's $d$) (**Figure 5a**), supporting the hypothesis that tumor regions are necessary for high ROR-P predictions.

To evaluate whether tumor tissue alone was sufficient for high ROR-P predictions, we applied HIPPO to generate synthetic slides containing only tumor regions, with all non-tumor tissue patches removed (**Figure 5b**). For seven models (i.e., CTransPath, DiRLv2, H-optimus-0, Kaiko-L/14, Prov-GigaPath, UNI, Virchow2), this manipulation did not result in a statistically significant change in the predicted probability of high ROR-P ($P > 0.05$, two-sided paired t-test), indicating that tumor regions alone were sufficient to reproduce the original high-risk predictions. Among the remaining models, all except Hibou-L exhibited significant increases in predicted probability following removal of non-tumor regions. However, effect sizes for most pathology foundation models were modest (Cohen's $d < 0.23$), further supporting the sufficiency of tumor morphology in driving high ROR-P predictions. The largest increase in predicted probability was observed for ResNet50 (Cohen's $d = 0.65$) and ViT-ResNet50 (Cohen's $d < 0.72$), suggesting that these non-pathology models were most affected by non-tumor tissue. Together, these perturbation-based analyses (**Figure 5a,b**) demonstrate that tumor regions are both necessary and, in most models, sufficient to drive high ROR-P predictions. These findings strengthen confidence that high-risk predictions of pathology foundation models reflect biologically relevant tumor morphology, rather than spurious associations with non-tumor components.

**Identifying candidate tissue biomarkers of high recurrence risk**

The combination of ABMIL models and explainable AI provides an opportunity not only to interpret model predictions, but also to discover morphology associated with recurrence risk. We developed a data-driven strategy to identify specific tissue regions that are sufficient to drive high-risk predictions. Informed by Kaelin's (2017) emphasis on biomarker sufficiency as a critical criterion in cancer research,[46] we leveraged HIPPO to automatically search for minimal regions of tissue that could convert a low- or medium-risk prediction into a high-risk one. This approach treats the trained model as a proxy observer, systematically identifying tissue regions that, when introduced into a low-risk specimen, consistently elevate predicted risk, revealing patterns that may serve as candidate, interpretable biomarkers of recurrence risk.

To operationalize this idea, we used HIPPO to perform a model-guided search for tissue patches sufficient to increase ROR-P predictions. Patches from high-risk WSIs were systematically inserted into low-risk WSIs to identify the smallest region capable of flipping the model's prediction (**Figure 5c**). This process yielded a set of 120 high-risk patches (~2.0 mm$^2$) that consistently elevated ROR-P scores across models and specimens. The identified regions exhibited features associated with aggressive tumor biology, including nuclear pleomorphism, mitotic activity, necrosis, and invasive growth, and there was a lack of tumor-infiltrating lymphocytes (**Figure 5d**, **Supplementary Figure 3**). To assess generalizability, these high-risk patches were introduced into additional low- and medium-risk slides. In all cases, the manipulated slides showed consistent increases in predicted ROR-P (**Figure 5e**, **Supplementary Figure 4**), suggesting that the identified regions are robust, transferable, and sufficient to drive high-risk predictions. These results demonstrate the potential of ABMIL models, in combination with explainable AI, to propose candidate histologic biomarkers of recurrence risk.

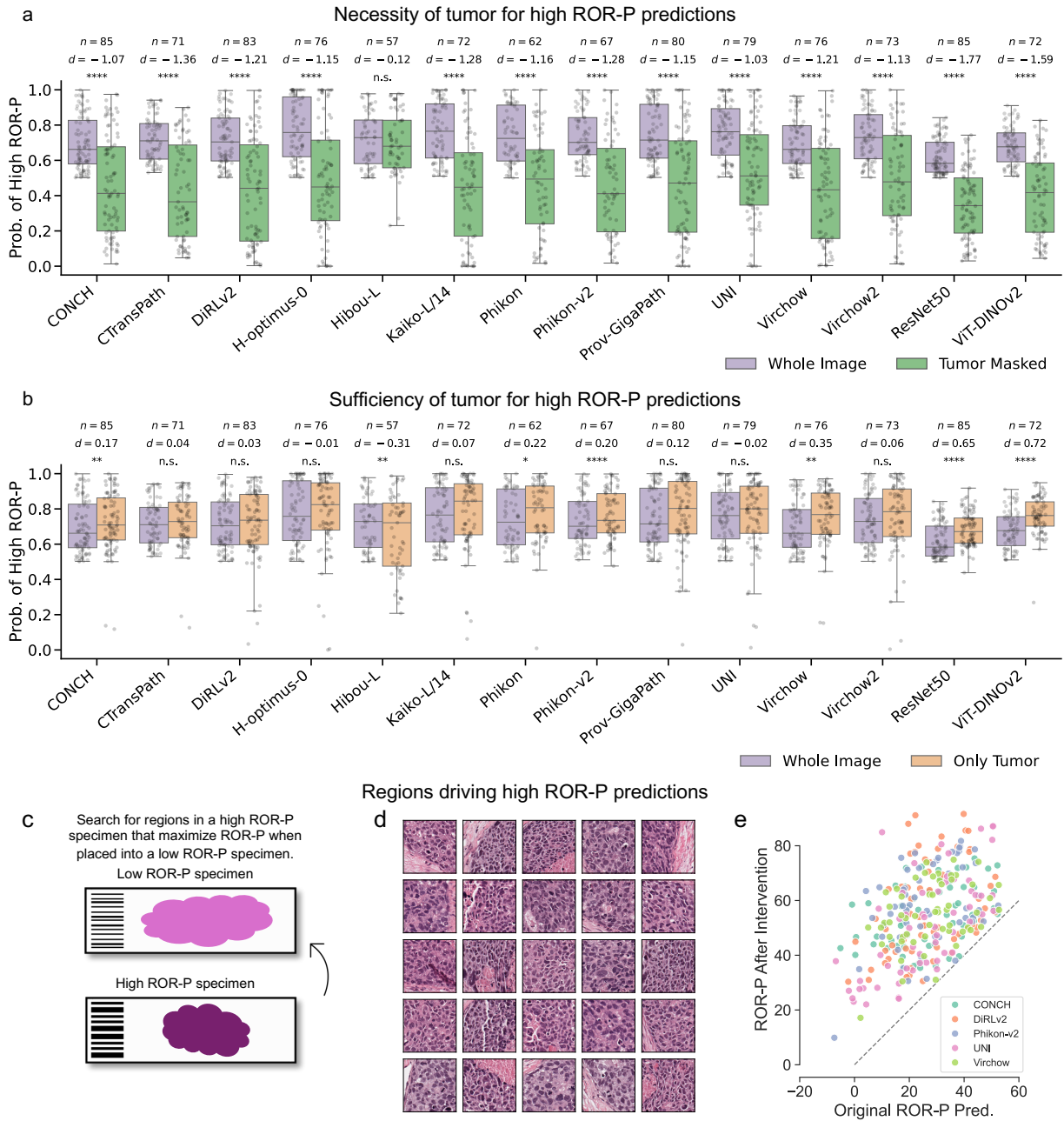

Figure 5. **Perturbation-based interpretability analyses. a,** Box plots showing the effect of loss of tumor patches on the predictions of high ROR-P using the classification model. Box plots show the first and third quartiles, the median (central line), and the range of data. Sample sizes and significance values are shown (*: $P < 0.05$, **: $P < 0.01$, ***: $P < 0.001$, ****: $P < 0.0001$, n.s.: $P > 0.05$; two-sided paired t-test). Sample sizes differ across models because only the specimens predicted as high ROR-P were used for each model. **b,** Box plots showing results of tumor sufficiency tests. The purple box plots are the same as the purple boxes in **a**, and orange boxes represent high ROR-P probabilities when non-tumor patches were removed. **c-e,** Search for regions that are sufficient to drive high ROR-P. **c,** Schematic of search strategy. Patches from high ROR-P specimens (n=5) were placed into low ROR-P specimens (n=5), and the image patch that resulted in the largest increase in predicted ROR-P was kept. This was continued until a stopping condition was met (Methods). **d,** Examples of the tissue patches sufficient to drive high ROR-P in an originally low-ROR-P specimen. **e,** Scatter plot of original ROR-P predictions and ROR-P predictions after the addition of 120 patches found via HIPPO search. The dashed gray line indicates the identify function. The full set of 120 patches is shown in Supplementary Figure 3, and the effects on other models are shown in Supplementary Figure 4.

## Discussion

In the present study, we demonstrate that computational pathology can accurately infer recurrence risk in breast cancer and systematically benchmark 12 pathology foundation models and two non-pathology baseline models using a unified framework, **MAKO**. Our results show that several foundation models, particularly CONCH, achieve robust performance in predicting both categorical and continuous PAM50-based ROR-P scores from H&E-stained WSIs. When evaluated for stratifying patients by recurrence events, these histology-based models performed comparably to transcriptomic assays. Beyond prediction, we show that pathology foundation models can be interrogated using perturbation-based methods to identify minimal tissue regions sufficient to drive high-risk predictions, suggesting their utility not only for histology-based risk stratification but also for biomarker discovery. Together, these findings highlight the promise of pathology foundation models as scalable, interpretable tools for the prediction of recurrence risk in breast cancer.

Our study builds upon a growing body of work demonstrating that deep learning models can infer molecular phenotypes and recurrence risk from histology. Couture et al., for example, showed that an ImageNet-pretrained convolutional neural network could be used to predict PAM50 molecular subtype, hormone receptor status, and ROR-PT (a score related to ROR-P that includes tumor size) from breast cancer tissue microarrays (TMAs).[23] We extend this work by using resections rather than TMAs, evaluating a diverse set of pathology-specific foundation models rather than a single ImageNet-based encoder, and introducing perturbation-based interpretation to assess the role of tumor regions in model predictions. More recently, Boehm et al. applied deep learning using the CTransPath pathology encoder to infer OncotypeDX® scores from histology specimens of early hormone receptor-positive breast cancer.[47] Our work builds on this by focusing specifically on the PAM50-based ROR-P score, benchmarking a broad set of pretrained pathology foundation models (including CTransPath), and incorporating survival analyses to directly assess model performance in predicting recurrence events. In addition, we demonstrate how perturbation-based methods can be used to identify necessary and sufficient regions for high-risk predictions and to uncover candidate histologic biomarkers of recurrence. Finally, El Nahhas et al. showed the value of regression-based deep learning for predicting continuous molecular biomarkers, such as homologous recombination deficiency.[43] We build on this by showing that regression-based ABMIL models can effectively predict continuous ROR-P scores and that perturbation experiments can be applied in this setting to quantify how specific tissue regions influence predicted continuous ROR-P scores. However, we also found that continuous predicted scores, when thresholded using the same thresholds as the transcriptomic ROR-P score, resulted in lower concordance indexes than predictions from models trained to directly predict risk categories. This suggests that the thresholds would have to be optimized for this image-based approach. Together, our study advances prior work by providing a standardized benchmarking framework, evaluating both classification and regression models, and introducing a rigorous, model-driven approach for biomarker discovery and interpretation.

This study has several notable strengths. We benchmarked 14 models using a large and diverse training cohort (CBCS) with long-term clinical follow-up and validated them on an independent external dataset (TCGA BRCA). By comparing pathology foundation models to general-purpose vision encoders, we highlight the added value of domain-specific pretraining for histopathology. We also demonstrate how ABMIL can be paired with perturbation-based interpretability to identify sufficient tissue regions for high-risk predictions, providing interpretable evidence of biologically grounded model behavior. Among the evaluated models,

CONCH emerged as a consistent top performer, achieving the highest classification performance across both datasets and ranking among the best models in regression tasks, suggesting that it may offer a particularly robust feature representation for recurrence risk modeling.

The primary limitation of the present study is that models were trained and evaluated on a cohort in which participants received heterogeneous treatments. As a result, we cannot determine whether the observed stratification of recurrence risk reflects prognostic information alone or is partially confounded by treatment effects. Unlike randomized clinical trials or studies with uniform treatment protocols, CBCS does not allow us to disentangle these factors. Therefore, while the predictions of our models align well with transcriptomic ROR-P (as measured by ROC AUC, Pearson r, and C-index), we cannot conclude that they would match the assay's performance in guiding treatment decisions or predicting outcomes in homogeneously treated populations. This limitation highlights the importance of evaluating histology-based models in randomized clinical trials. Additionally, while external validation on TCGA BRCA supports generalizability, performance was variable, potentially due to differences in cohort characteristics and slide preparation. Finally, although HIPPO enables localized perturbation to test model reasoning, it reflects causality as inferred by the model rather than true biological mechanisms.

Our work opens several avenues for future research. Prospective studies in clinical settings are essential to assess real-world performance and determine the clinical utility of histology-based recurrence prediction. Integrating these models with additional clinical, genomic, or spatial transcriptomic data may further improve accuracy and interpretability. The perturbation-based approach presented here could also be extended to other prediction tasks or used to guide targeted biomarker discovery. As foundation models continue to evolve, benchmarking efforts like MAKO will be critical for identifying performant, generalizable models and for ensuring that model outputs are interpretable, biologically meaningful, and clinically actionable.

## Methods

**Study population**

This study involved data from The Carolina Breast Cancer Study (CBCS), which has been described previously.[48,49] Briefly, CBCS is a multidisciplinary study of invasive breast cancer that enrolled a total of 2,998 female participants ages 20-74 years from 44 counties in North Carolina. CBCS oversampled self-identified Black/African American women and younger women (age < 50 years). Cases were identified by rapid case ascertainment via the UNC Rapid Case Ascertainment Core in conjunction with North Carolina Central Cancer Registry (diagnosis years 2008-2013). The current analysis includes the subset of participants with digitized whole slide images (WSIs) and corresponding ROR-P scores (n=1,339). Furthermore, we restricted to the 63% (n=847) of these participants with ER-positive, HER2-negative disease and pathologic stages I, II, or III (**Table 1**). The samples with WSIs had similar distributions of age, race, stage, size, and node status as the overall CBCS3 study population. Details regarding formalin-fixed paraffin-embedded (FFPE) and immunohistochemistry (IHC) preparation have been described previously.[50]

As a validation set, we included data from the invasive breast cancer cohort of The Cancer Genome Atlas (TCGA BRCA).[42] A total of 1,050 participants were analyzed, of whom 62% (n=628) had ER-positive, HER2-negative and stage I–III tumors.

**Molecular scoring**

The methods for computing PAM50 centroid correlation coefficients, intrinsic subtypes, and the risk of recurrence (ROR-P) score have been described previously.[48] Briefly, tissue cores $1.0\ mm$ in diameter were sampled from tumor-rich regions (these cores contain significantly less tissue than the full whole slide images used for training neural networks in this study). Bulk RNA counting via NanoString nCounter was performed to derive PAM50 centroid correlation coefficients, which quantify the similarity of a tumor's gene expression profile to each PAM50 subtype.[51] The assigned PAM50 subtype corresponds to the subtype with the highest correlation coefficient. The ROR-P score was computed as a weighted sum of the PAM50 correlation coefficients and a proliferation-related component per the PAM50 algorithm in Parker et al.[12] The distribution of ROR-P scores in CBCS and TCGA BRCA were slightly different, with TCGA BRCA showing lower ROR-P scores on average (Supplementary Figure 1).

**Whole slide image processing**

Formalin-fixed paraffin-embedded (FFPE) tissue specimens in CBCS were scanned using an Aperio scanner (Leica Biosystems, Nussloch, Germany) at $20 \times$ magnification (approximately $0.50\ \frac{\mu m}{pixel}$, MPP). WSIs from TCGA BRCA were downloaded from the Genomic Data Commons Data Portal. Seven slides from TCGA BRCA were excluded because the metadata specifying the physical size of each pixel (MPP) was missing. Patch coordinates were calculated using the CLAM toolkit,[39] which was modified to create patches of a constant physical size. Tissue image patches of size $128 \times 128\ \mu m^2$ were extracted, and the same patch coordinates were used for all models trained in the present report.

Each patch was embedded using pre-trained feature extraction models. We evaluated 12 foundation models trained on pathology images: CONCH,[32] CTransPath,[52] DiRLV2,[53] Hibou-L,[34] H-optimus-0,[33] Kaiko-L/14,[54] Phikon,[55] Phikon-v2,[56] Prov-GigaPath,[35] UNI,[31] Virchow,[36] and Virchow2.[57] As a comparison, we also embedded patches using ResNet50,[58] which was trained on ImageNet, and ViT-DINOv2,[59] which was trained on over 142 million natural images. For each WSI, all patches were embedded using one model at a time, with each patch transformed into a corresponding feature vector. The resulting vectors from all patches within a WSI were concatenated to form a WSI-level matrix corresponding to that feature extraction model. This process was repeated separately for each of the 14 feature extraction models. The resulting matrices were then used as inputs to the WSI-level models.

**WSI-level neural network modeling**

Attention-based multiple instance learning (ABMIL) was used to learn WSI-level labels from patch embeddings.[38] The ABMIL models first encoded patch-level feature vectors using a fully connected layer ($L = 512$), rectified linear unit activation, and dropout of $p = 0.25$ for regularization. A gated attention mechanism then computed attention scores by applying parallel tanh-activated and sigmoid-activated branches (dimensionality $D = 384$), followed by element-wise multiplication and a linear projection. The resulting attention scores were normalized via softmax, and the attention-weighted sum of patch embeddings formed a WSI-level feature vector, which was then processed by a final linear layer for classification or regression. The models output two logits for risk classification and one for ROR-P regression.

We employed a 10-fold cross-validation procedure to assess the generalization performance of our models. Specifically, our cohort of 1,339 CBCS participants was divided into 10 equally sized subsets (folds). The subsets were stratified such that the distribution of ROR-P groups was similar across folds. Iteratively, each fold served exactly once as the test set, while the remaining nine folds formed a combined dataset that was subsequently partitioned into separate training and validation subsets. For each iteration, a model was trained using only the training subset, optimized using predictions evaluated exclusively on the validation subset, and finally assessed using predictions generated on the independent test fold. After completing all 10 iterations, we obtained a complete set of out-of-sample predictions for the entire dataset. Critically, each participant appeared in only one partition per iteration, ensuring that the model was never trained or optimized on data from participants included in the corresponding test set. This procedure preserved the integrity of the training-validation-testing separation and prevented data leakage, thus providing an unbiased evaluation of model performance.

**Model evaluation on TCGA BRCA**

Models trained on CBCS were applied to TCGA BRCA slides using the same pipeline. Patch embeddings from TCGA BRCA served as inputs to ABMIL models trained on CBCS. Because CBCS models were developed using 10-fold cross-validation, inference was performed with all 10 models. The final prediction for each TCGA BRCA specimen was obtained by averaging the softmax-transformed logits across the 10 models.

**Statistical analyses**

For classification of ROR-P risk categories, we evaluated model performance using the area under the receiver operating characteristic curve (ROC AUC), and DeLong's test was used to assess statistical significance of differences in model performance.[60] For continuous ROR-P prediction tasks, model performance was evaluated using Pearson correlation, and the statistical significance of differences in Pearson correlation coefficients was evaluated using Meng's z-test for comparing dependent correlations.[61] Pairwise comparisons were performed between each foundation model and the ResNet50 baseline. To correct for multiple hypothesis testing, p-values were adjusted using the Benjamini-Hochberg procedure (FDR).[62] All statistical tests were two-sided. Analyses were conducted separately for the CBCS and TCGA BRCA cohorts.

We assessed model performance for recurrence-free survival using time-to-event data. Survival time was defined as the number of years from diagnosis to the first recurrence or censoring at 10 years. The concordance index (C-index) was used to quantify the discriminative ability of each model. C-indices were computed using the concordance() function from the "survival" R package. C-index values range from 0.5 (no discrimination) to 1.0 (perfect discrimination). For comparison between model predictions and the reference transcriptomic score, we used the "compare" R package, which implements a statistical test for comparing correlated C-indices.[44] P values were adjusted for multiple comparisons using the FDR correction.[62] To assess whether model predictions were significantly associated with recurrence-free survival, we fit univariable Cox proportional hazards models using the coxph() function from the "survival" R package. The continuous model prediction score was used as the sole predictor. We also applied the log-rank test using the survdiff() function implemented in the "survminer" R package.

For categorical prediction models, model logits were transformed with softmax and then converted to binary risk classifications using optimized thresholds. Rather than applying a fixed threshold (e.g., 0.5), we determined an optimal threshold for each encoder by maximizing Youden's J ($J = sensitivity + specificity - 1$).[63] Specifically, we performed 10-fold cross-validation, and for each model, we concatenated the validation set predictions for ER-positive, HER2-negative specimens across all folds. Youden's J was then computed on this pooled validation set to identify the optimal threshold, which was subsequently applied to the model's test set predictions. This procedure was performed independently for each model. This approach ensured that threshold selection reflected optimal risk discrimination rather than arbitrary decision boundaries, thus providing a more meaningful evaluation of clinical utility.

**Interpretation of ABMIL models**

Attention maps from ABMIL models were used as a preliminary tool for qualitative interpretation. These maps were generated in Python, converted to GeoJSON format (with each patch polygon annotated with its ranked attention score), and subsequently visualized in QuPath[64] for assessment.

**HIPPO experiments**

To further interpret model predictions, we applied HIPPO, a quantitative, occlusion-based explainability technique.[40] As the ROR-P was developed using tumor-intrinsic genes, we hypothesized that tumor regions are necessary and sufficient for model predictions of ROR-P. We focused on ROR-P classification models. Tumor masks were generated using PenAnnotationExtractor,[65] which extracted tissue regions based on pen annotations on the glass slide. These masks defined the tumor-containing areas within each WSI. To evaluate the effect of tumor removal on high ROR-P predictions, we excluded all patch embeddings that intersected with the tumor mask and compared the model's softmax probabilities between the original and tumor-removed WSIs. This analysis was restricted to slides where the model originally predicted high ROR-P with a softmax probability greater than 0.5.

To assess the sufficiency of tumor regions for ROR-P predictions, we removed all patches that did not intersect with the tumor mask, retaining only tumor-containing regions. As in the necessity analysis, we focused on WSIs where the models originally predicted high ROR-P with a softmax probability greater than 0.5. We then compared the model's predictions between the original WSIs and those containing only tumor regions.

Beyond the hypothesis-driven tests described above, we conducted data-driven experiments to identify regions sufficient to flip a low ROR-P specimen to high ROR-P. This approach aimed to uncover robust tissue biomarkers that the models associated with recurrence risk. We selected five high ROR-P slides and five low ROR-P slides and applied a greedy search strategy using the ROR-P regression model trained with the h-Optimus-0 features. In each iteration, a single patch from a high ROR-P slide was inserted into a low ROR-P slide, and the model's prediction was re-evaluated. The patch was then replaced with the next patch from the high ROR-P slide, and this process was repeated for all patches. The patch that produced the highest increase in ROR-P when inserted into the low ROR-P slide was retained, and the search continued with the remaining patches. This iterative process was performed until either a quarter of the patches in the high ROR-P specimen were used or the manipulated specimen's predicted ROR-P reached the high-risk threshold.

To assess the robustness and generalizability of the identified high ROR-P patches, we inserted them into additional low- and medium-ROR-P specimens and measured the resulting changes in predicted ROR-P scores. This analysis was conducted across all foundation models to evaluate the robustness of learned risk-associated features.

## Data availability

Carolina Breast Cancer Study is actively following participants and under an IRB-approved protocol that does not permit data sharing on public websites. However, the study shares data through an IRB-approved data use agreement system as described on its website (https://unclineberger.org/cbcs/). The results shown here are in whole or part based upon data generated by the TCGA Research Network: https://www.cancer.gov/tcga.

## Code availability

Code to reproduce the analyses presented in this manuscript is available at https://github.com/kaczmarj/MAKO.

## Acknowledgements

This research was supported awards from the National Cancer Institute (5UH3CA225021, U24CA215109), National Science Foundation (IIS-2212046), the Stony Brook University Provost's ProFund, support from the Medical Scientist Training Program at Stony Brook University (T32GM008444), and generous donor support from Bob Beals and Betsy Barton. This research was also supported by a grant from UNC Lineberger Comprehensive Cancer Center, which is funded by the University Cancer Research Fund of North Carolina, the Susan G Komen Foundation (OGUNC1202, OG22873776, SAC210102, TREND21686258), National Cancer Institute (R01CA253450, U24CA264021), the National Cancer Institute Specialized Program of Research Excellence (SPORE) in Breast Cancer (NIH/NCI P50-CA058223), the Breast Cancer Research Foundation (HEI-23-003), and the US Department of Defense (HT94252310235). This research was funded, in part, by the Advanced Research Projects Agency for Health (ARPA-H) under contract 140D042490008. The views and conclusions contained in this document are those of the authors and should not be interpreted as representing the official policies, either expressed or implied, of the U.S. Government. This research recruited participants &/or obtained data with the assistance of Rapid Case Ascertainment, a collaboration between the North Carolina Central Cancer Registry and UNC Lineberger. RCA is supported by a grant from the National Cancer Institute of the National Institutes of Health (P30CA016086). The authors would like to acknowledge the University of North Carolina BioSpecimen Processing Facility for sample processing, storage, and sample disbursements (http://bsp.web.unc.edu/) and the Breast Cancer Research Foundation HEI-23-003. We are grateful to CBCS participants and study staff.

## Author contributions

JRK, MAT, KAH, and JHS conceived the project and designed the experiments. SCVA generated the PAM50 and ROR-P results for CBCS. JRK developed the software, trained the models, conducted all computational analyses, and performed statistical tests. JRK and PKK designed the HIPPO analyses. JRK, PKK, and AJC interpreted results of the HIPPO experiments. AJC and JRK interpreted the attention heatmaps. RG provided critical feedback. JRK drafted the initial manuscript. All authors contributed to manuscript revisions and approved the final version. KAH and JHS jointly supervised the work.

## Competing interests

The University of North Carolina, Chapel Hill has a license of intellectual property interest in GeneCentric Diagnostics and BioClassifier, LLC, which may be used in this study. The University of North Carolina, Chapel Hill may benefit from this interest that is/are related to this research. The terms of this arrangement have been reviewed and approved by the University of North Carolina, Chapel Hill Conflict of Interest Program in accordance with its conflict of interest

# Supplementary Information

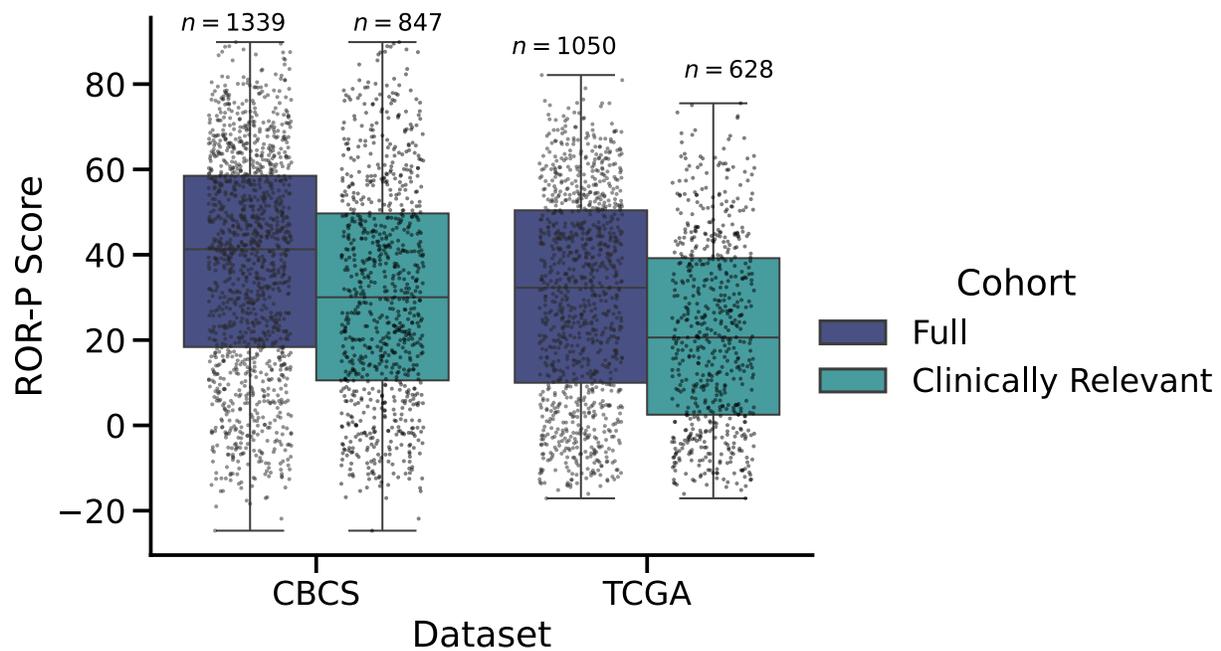

Supplementary Figure 1. Distribution of transcriptomic ROR-P scores in CBCS and TCGA BRCA cohorts. Box plots illustrate ROR-P score distributions in CBCS ($n = 1,339$ participants) and TCGA BRCA ($n = 1,050$ participants) datasets, stratified by cohort type (full dataset vs. clinically relevant subset). The clinically relevant subset includes only participants with ER-positive, HER2-negative, non-metastatic breast cancer (the target population for the ROR-P assay). Box plots show the first and third quartiles, the median (central line), and the range of data (whiskers). Individual data points represent discrete participants.

Supplementary Table 1. Performance of ROR-P classification. N corresponds to the number of WSIs in the analysis. ROC AUC denotes the receiver operating characteristic area under curve. PR AUC denotes the precision-recall area under curve. The P values shown were calculated using DeLong's test and were adjusted for multiple comparisons (FDR) within each dataset. Rows are sorted in descending order of ROC AUC in CBCS.

| | CBCS | | | | TCGA | | | |
|---|---|---|---|---|---|---|---|---|
| | N | ROC AUC | PR AUC | Adj. P | N | ROC AUC | PR AUC | Adj. P |
| CONCH | 883 | 0.809 | 0.589 | $4.47 \times 10^{-4}$ | 628 | 0.852 | 0.461 | $2.76 \times 10^{-2}$ |
| UNI | 883 | 0.808 | 0.541 | $3.29 \times 10^{-3}$ | 628 | 0.825 | 0.355 | $7.34 \times 10^{-2}$ |
| Phikon | 883 | 0.799 | 0.534 | $2.85 \times 10^{-3}$ | 628 | 0.824 | 0.289 | $7.34 \times 10^{-2}$ |
| CTransPath | 883 | 0.797 | 0.556 | $4.47 \times 10^{-4}$ | 628 | 0.829 | 0.446 | $2.76 \times 10^{-2}$ |
| Phikon-v2 | 883 | 0.791 | 0.558 | $1.21 \times 10^{-2}$ | 628 | 0.829 | 0.448 | $7.34 \times 10^{-2}$ |
| Virchow | 883 | 0.790 | 0.549 | $1.42 \times 10^{-2}$ | 628 | 0.780 | 0.257 | $7.90 \times 10^{-1}$ |
| Kaiko-L/14 | 883 | 0.786 | 0.530 | $5.42 \times 10^{-2}$ | 628 | 0.779 | 0.286 | $7.90 \times 10^{-1}$ |
| Prov-GigaPath | 883 | 0.784 | 0.513 | $6.80 \times 10^{-2}$ | 628 | 0.824 | 0.380 | $7.34 \times 10^{-2}$ |
| DiRLv2 | 883 | 0.783 | 0.512 | $1.42 \times 10^{-2}$ | 628 | 0.794 | 0.396 | $3.90 \times 10^{-1}$ |
| Hibou-L | 883 | 0.774 | 0.549 | $8.96 \times 10^{-2}$ | 628 | 0.745 | 0.311 | $5.54 \times 10^{-1}$ |
| Virchow2 | 883 | 0.769 | 0.542 | $2.12 \times 10^{-1}$ | 628 | 0.842 | 0.452 | $5.06 \times 10^{-2}$ |
| H-optimus-0 | 883 | 0.763 | 0.490 | $3.71 \times 10^{-1}$ | 628 | 0.841 | 0.455 | $5.06 \times 10^{-2}$ |
| ViT-DINOv2 | 883 | 0.762 | 0.513 | $1.82 \times 10^{-1}$ | 628 | 0.803 | 0.323 | $1.43 \times 10^{-1}$ |
| ResNet50 | 883 | 0.745 | 0.488 | Reference | 628 | 0.772 | 0.263 | Reference |

Supplementary Table 2. Performance of ROR-P regression. N corresponds to the number of WSIs in the analysis. PCC denotes the Pearson correlation coefficient (r). The P values shown are calculated using Meng's test relative to the PCC of ResNet5. Rows are sorted in order of descending PCC in CBCS.

|  | CBCS | | | TCGA | | |
|---|---|---|---|---|---|---|
|  | N | PCC | Adj. P | N | PCC | Adj. P |
| H-optimus-0 | 883 | 0.6375150 | $2.94 \times 10^{-6}$ | 628 | 0.4428270 | $6.37 \times 10^{-2}$ |
| Prov-GigaPath | 883 | 0.6351038 | $2.05 \times 10^{-6}$ | 628 | 0.5603631 | $1.74 \times 10^{-1}$ |
| Virchow | 883 | 0.6295329 | $2.94 \times 10^{-6}$ | 628 | 0.5263878 | $8.85 \times 10^{-1}$ |
| Virchow2 | 883 | 0.6267673 | $9.52 \times 10^{-6}$ | 628 | 0.5865176 | $5.28 \times 10^{-2}$ |
| CONCH | 883 | 0.6242045 | $1.23 \times 10^{-5}$ | 628 | 0.5606724 | $1.74 \times 10^{-1}$ |
| UNI | 883 | 0.6223194 | $4.89 \times 10^{-5}$ | 628 | 0.5192497 | $9.70 \times 10^{-1}$ |
| Phikon-v2 | 883 | 0.6193370 | $5.93 \times 10^{-5}$ | 628 | 0.4946317 | $5.20 \times 10^{-1}$ |
| CTransPath | 883 | 0.5973843 | $1.07 \times 10^{-3}$ | 628 | 0.5634536 | $5.81 \times 10^{-2}$ |
| Kaiko-L/14 | 883 | 0.5962158 | $6.61 \times 10^{-3}$ | 628 | 0.4261795 | $5.28 \times 10^{-2}$ |
| Phikon | 883 | 0.5843723 | $4.25 \times 10^{-2}$ | 628 | 0.4564143 | $1.17 \times 10^{-1}$ |
| Hibou-L | 883 | 0.5809827 | $4.45 \times 10^{-2}$ | 628 | 0.5149466 | $9.70 \times 10^{-1}$ |
| DiRLv2 | 883 | 0.5705349 | $1.05 \times 10^{-1}$ | 628 | 0.4966830 | $4.71 \times 10^{-1}$ |
| ResNet50 | 883 | 0.5405439 | Reference | 628 | 0.5182527 | Reference |
| ViT-DINOv2 | 883 | 0.5183641 | $2.12 \times 10^{-1}$ | 628 | 0.5523207 | $1.33 \times 10^{-1}$ |

Supplementary Table 3. Thresholds for ROR-P classification optimized using Youden's J. Thresholds were determined using validation data, which is distinct from the test sets used to report performance metrics in this study.

| Encoder | Threshold |
| --- | --- |
| CHIEF | 0.2368 |
| CONCH | 0.3050 |
| CTransPath | 0.2595 |
| DiRLv2 | 0.2760 |
| GigaPath | 0.3177 |
| Hibou | 0.1870 |
| Kaiko | 0.2188 |
| Optimus | 0.1555 |
| Phikon | 0.2723 |
| Phikon-v2 | 0.2868 |
| REMEDIS | 0.2936 |
| ResNet50 | 0.3331 |
| UNI | 0.2092 |
| ViT-DINOv2 | 0.2614 |
| Virchow | 0.1985 |

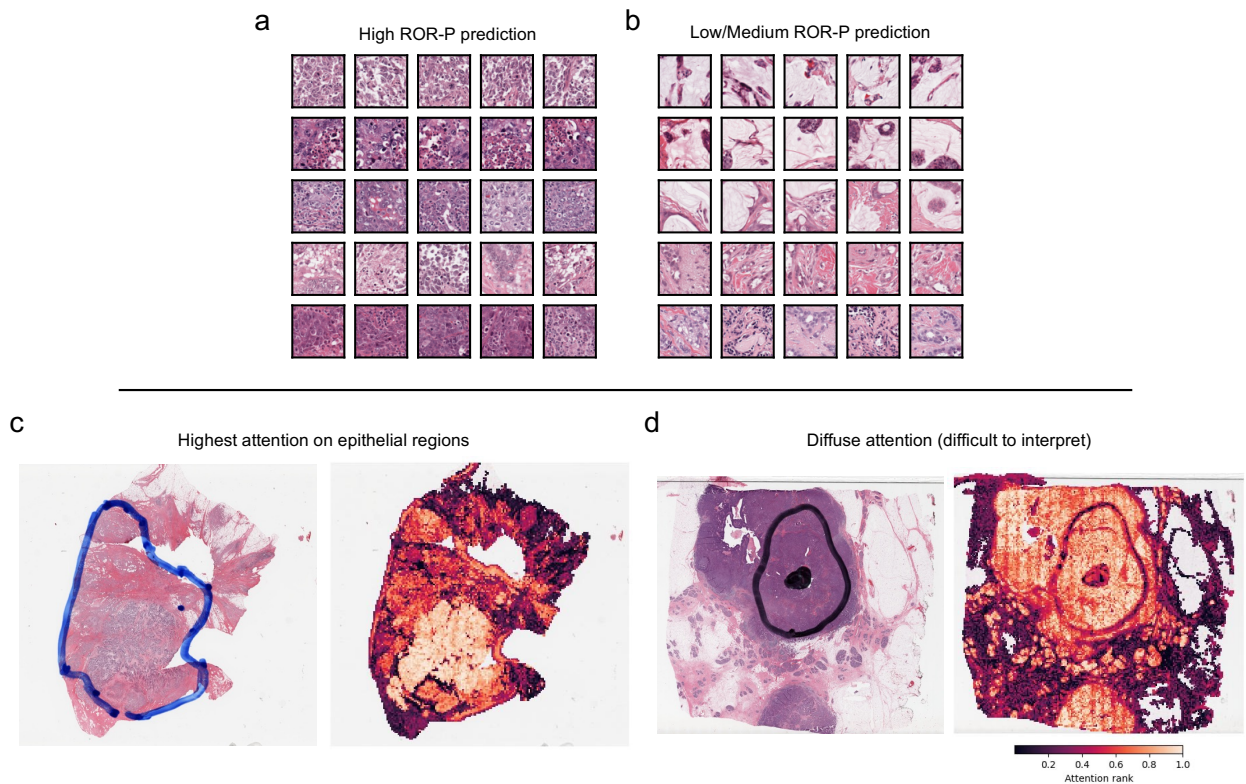

Supplementary Figure 2. Visualization of attention weights from ABMIL models predicting binarized ROR-P. **a-b**, Representative image patches with the highest attention weights identified by ABMIL models as (**a**) high ROR-P and (**b**) low/medium ROR-P. Each row represents a distinct whole slide image (WSI). **c-d**, Example WSIs demonstrating interpretability of attention maps in WSIs predicted as high ROR-P (left: WSI, right: corresponding attention heatmap). **c**, Attention heatmap tends to highlight epithelial regions, as noted by lighter shades. **d**, A contrasting example highlighting a diffuse and less interpretable attention heatmap. This emphasizes the limitations of attention maps alone for interpretation of ABMIL models.

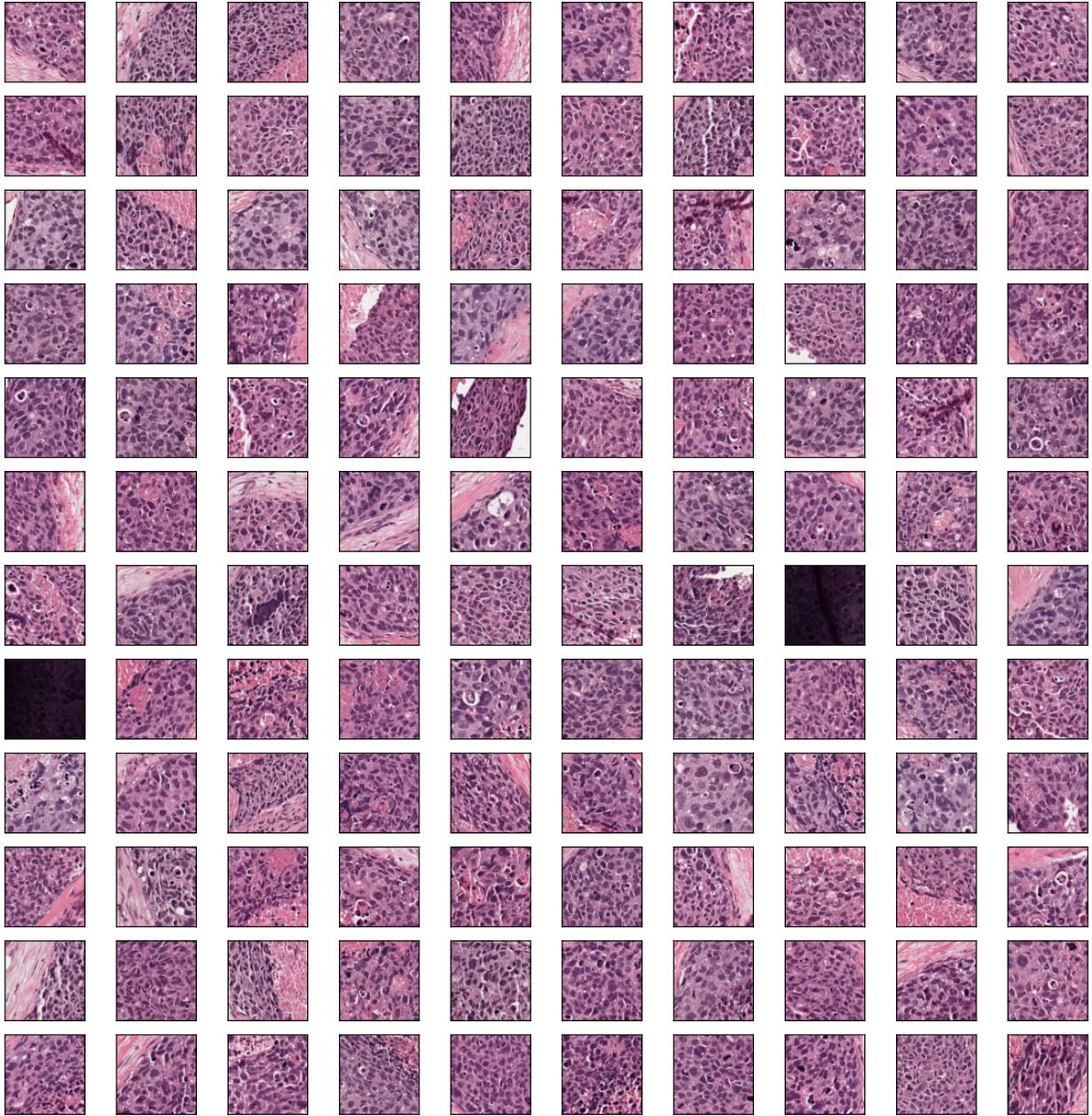

Supplementary Figure 3. Tissue regions sufficient to increase ROR-P predictions. These patches were identified using HIPPO greedy search that selected regions that maximized ROR-P model predictions when inserted into a low- or medium-ROR-P WSI. All patches were derived from a single slide and exhibit key histologic features, including invasive margins, regions of necrosis, pronounced nuclear pleomorphism, and numerous mitotic figures. Notably, tumor-infiltrating lymphocytes are sparse in these regions. Two dark regions are present, representing patches containing marker.

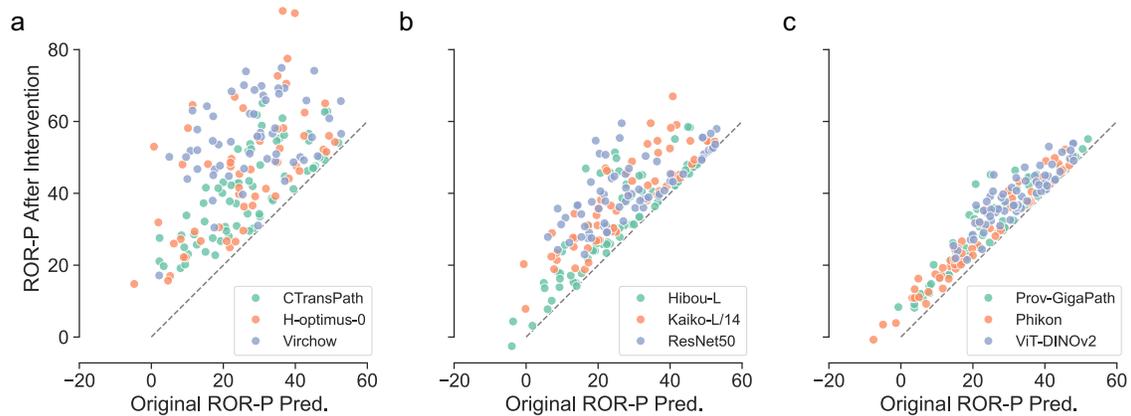

Supplementary Figure 4. Scatter plots of original ROR-P predictions and ROR-P predictions after the addition of 120 patches found via HIPPO search. The dashed gray line indicates the identify function. Within each panel, each point represents a single whole slide image, and only whole slide images that were predicted as low/medium ROR-P were included. Models are represented in different subpanels for ease of viewing: **a**, CTransPath, H-optimus-0, and Virchow; **b**, Hibou-L, Kaiko-L/14, ResNet50; **c**, Prov-GigaPath, Phikon, ViT-DINOv2.